\documentstyle[11pt]{article}
\newtheorem{theorem}{Theorem}

\newtheorem{corollary}{Corollary}
\newcommand {\dfn} {\stackrel{\Delta} {=}}

\newcommand {\reals} {{\rm I\!R}}

\newcommand {\bu} {\mbox{\boldmath $u$}}
\newcommand {\bv} {\mbox{\boldmath $v$}}

\newcommand {\bx} {\mbox{\boldmath $x$}}
\newcommand {\by} {\mbox{\boldmath $y$}}
\newcommand {\bz} {\mbox{\boldmath $z$}}

\newcommand {\bX} {\mbox{\boldmath $X$}}
\newcommand {\bY} {\mbox{\boldmath $Y$}}

\newcommand{\calC}{{\cal C}}
\newcommand{\calD}{{\cal D}}
\newcommand{\calE}{{\cal E}}
\newcommand{\calF}{{\cal F}}
\newcommand{\calG}{{\cal G}}

\newcommand{\calS}{{\cal S}}
\newcommand{\calT}{{\cal T}}
\newcommand{\calU}{{\cal U}}
\newcommand{\calV}{{\cal V}}

\newcommand{\calX}{{\cal X}}
\newcommand{\calY}{{\cal Y}}

\newcommand{\hu}{{\hat{u}}}
\newcommand{\hx}{{\hat{x}}}

\newcommand{\hX}{{\hat{X}}}

\newcommand{\tx}{{\tilde{x}}}
\newcommand{\tX}{{\tilde{X}}}

\begin{document}
\thispagestyle{empty}
\setcounter{page}{1}
\setlength{\baselineskip}{1.5\baselineskip}
\title{On the Wyner--Ziv Problem for Individual Sequences}
\author{Neri Merhav 
and Jacob Ziv
\\ \\
Department of Electrical Engineering \\
Technion - Israel Institute of Technology \\
Haifa 32000, ISRAEL}
\maketitle

\begin{abstract}
We consider a variation of the Wyner--Ziv 
problem pertaining to lossy compression
of individual sequences using finite--state encoders and decoders.
There are two main results in this paper. The first characterizes the
relationship between the performance 
of the best $M$--state encoder--decoder
pair to that of the best block code of size $\ell$ for every input sequence, 
and shows that the loss of the latter 
relative to the former (in terms of both 
rate and distortion) never exceeds the order of $(\log M)/\ell$,
independently of the input sequence.
Thus, in the limit of large $M$, the 
best rate--distortion performance of every infinite
source sequence can be approached universally 
by a sequence of block codes (which are also
implementable by finite--state machines). 
While this result assumes an asymptotic regime
where the number of states is fixed,
and only the length $n$ of the input sequence grows without
bound, we then consider the case where the
number of states $M=M_n$ is allowed to grow concurrently with $n$.
Our second result is then about the critical growth rate of $M_n$ such
that the rate--distortion performance of 
$M_n$--state encoder--decoder pairs can still be matched
by a universal code. We show that this critical 
growth rate is of $M_n$ is linear in $n$.

\vspace{1cm}

\end{abstract}

\section{Introduction}

In a series of papers from the late 
seventies until the mid--eighties, Ziv
\cite{Ziv78},\cite{Ziv80},\cite{Ziv84}, and Ziv and Lempel \cite{ZL78},\cite{LZ86}
have developed a theory of universal compression of
individual sequences using finite--state machines (FSM's). 
In particular, the work \cite{Ziv78}
focuses on universal, fixed--rate, (almost) 
lossless compression of individual sequences
using finite--state encoders and decoders, 
which was then further developed to the
well--known Lempel--Ziv algorithm \cite{ZL78},\cite{LZ86}.
In \cite{Ziv80}, the framework of \cite{Ziv78} was extended to
lossy compression, and in \cite{Ziv84}, the results 
of \cite{Ziv78} were extended in another
direction, pertaining to almost lossless compression 
in the presence of an (individual) side
information sequence at the decoder, namely, 
an analogue to Slepian--Wolf coding \cite{SW73} 
for individual sequences.

In this work, we take yet another step in this 
direction and further generalize this model setting,
of universal coding for individual sequences using
finite--state encoders and decoders, to that of
lossy compression in the presence of side 
information at the decoder, in other words,
Wyner--Ziv (W--Z) coding \cite{WZ76} for individual sequences. 
Also, unlike the fixed--rate codes assumed in
\cite{Ziv78},\cite{Ziv80},\cite{Ziv84}, here our model allows variable--rate
coding, which give rise to considerably more flexibility. 
On the other hand, in our model,
the side information sequence at the decoder
is assumed to be generated from 
the individual source sequence (to be compressed) via a known
memoryless channel, in contrast to \cite{Ziv84}, 
where it is modelled as another individual sequence.\footnote{The reason for
this assumption
is that even in the classical, probabilistic setting,
W--Z coding cannot be universal w.r.t.\ the channel from the
source to the side information stream
(unless there is feedback) as the encoder, which has no access to the
side information, cannot `learn' its statistics. This is different from
the Slepian--Wolf setting, where the encoder does not depend on these
statistics.} Furthermore, our model
setting can also be viewed as an 
extension of the setting of universal finite--state denoising
of individual sequences corrupted by stochastic noise (cf.\ \cite{WOSVW05}, 
\cite{OWWSM04} and references therein): The denoising problem is actually
a special case of this W--Z model, where the coding rate is zero.

There are two main results in this paper. The first result 
is a characterization of the
relationship between the 
performance of the best $M$--state finite--state encoder--decoder
pair (for a given input sequence) 
to that of the best block code of size $\ell$ for every input sequence,
and it shows that the loss of the latter
relative to the former (in terms of both rate and distortion) 
is of the order of $(\log M)/\ell$,
independently of the input sequence.
Thus, in the limit of large $M$, the best 
rate--distortion performance of every infinite
source sequence can be approached universally 
by a sequence of block codes (which are also
implementable by finite--state machines). One of the interesting features
of these universal codes is that they require no binning, as opposed to
the well--known W--Z code in classical in the probabilistic case \cite{WZ76}. 
We also extend this result to
framework of successive refinement coding 
(cf.\ e.g., \cite{EC91}, \cite{Rimoldi94}, \cite{SM04}),
where there are two encoders and two decoders (all of which are 
finite--state machines):
The first encoder transmits a relatively 
coarse description to the first decoder,
which has access also to a certain side information stream. The second encoder sends
a refinement code, to another decoder that has access also to the first compressed bitstream,
as well as to another side information sequence. This is setup is in analogy
to a recent study on successive refinement coding for the W--Z
problem for probabilistic memoryless sources \cite{SM04}.
However, in contrast to \cite{SM04}, where a certain Markov
structure had to be assumed regarding the source 
and the two side information processes, here no such structure is needed.
Also, unlike in \cite{SM04}, here the extension to multiple stages is straghtforward.

Returning to the single--stage coding model, we next relax
the assumption that the number of states is fixed, independently of
the length $n$ of the input data. In other words,
we examine an asymptotic regime that allows the number of states 
$M=M_n$ to grow concurrently with $n$, and we investigate
the critical growth rate of $M_n$ such
that the rate--distortion performance of 
$M_n$--state encoder--decoder pairs can still be matched
by a universal code (in a sense to be made precise later on).
Our second result is that this critical growth rate of $M_n$ is linear in $n$, in the sense
that if $M_n=n^\theta$, universal achievability is guaranteed for all $\theta < 1$, but
not for $\theta > 1$. In other words, $\theta=1$ is the critical value of $\theta$.

In this context, it is interesting to go back, for a moment, to the lossless case without
side information and to consider the performance of the well--known Lempel--Ziv algorithm in
that respect. By examining the converse to the coding theorem (Theorem 1) in \cite{ZL78},
which states that the best compression achievable by an FSM with $M$ states is lower
bounded by a quantity the behaves roughly like $c\log (c/4M^2)$ where $c$ is the number of
distinct phrases in the input sequence.
Since the length of the LZ code is about $c\log c$, the gap, $c\log(4M^2)$, would
be relatively negligible only as long as $\log M$ would 
be very small compared to $\log c \sim \O(\log n)$, which
is guarateed to be the case when $\theta$ is very small ($\theta << 1$). 
It, therefore, turns out that there is
a gap between the best that can be done by a universal code, where $\theta$ can be
chosen arbitrarily clode to unity, and the performance of the LZ algorithm in that respect.

The outline of this paper is as follows. In Section 2, we define
the problem and establish notation conventions. Section 3 is devoted
to the derivation of the first result described above, pertaining to a
fixed number of states. Finally, Section 4 focuses on the critical growth
rate of the number of states. The extension of the results of Section 3
to successive refinement coding is deferred to the Appendix, for the sake
of continuity between Sections 3 and 4 which are both about single--stage codes.

\section{Notation and Problem Formuation}

Throughout the paper, random variables will be denoted by capital
letters, specific values they may take will be denoted by the
corresponding lower case letters, and their alphabets, as well as some
other sets, will be denoted by calligraphic letters. Similarly, random
vectors, their realizations, and their alphabets, will be denoted,
respectively, by capital letters, the corresponding lower case letters,
and calligraphic letters, all superscripted by their dimensions. For
example, the random vector $Y^n=(Y_1,\ldots,Y_n)$, ($n$ -- positive
integer) may take a specific vector value $y^n=(y_1,\ldots,y_n)$
in $\calY^n$, the $n$th order Cartesian power of $\calY$, which is 
the alphabet of each component of this vector. For $i\le j$
($i$, $j$ -- positive integers), $x_i^j$ will denote the segment
$(x_i,\ldots,x_j)$, where for $i=1$ the subscript will be omitted.

Let $\bx=(x_1,x_2,\ldots)$, 
$x_i\in\calX$, $i=1,2,\ldots$, with $|\calX|=\alpha<\infty$,
be an infinite input sequence (to be compressed) and let $\by=(y_1,y_2,\ldots)$, $y_i\in\calY$, 
$i=1,2,\ldots$, with $|\calY|=\beta<\infty$, be a 
corresponding side--information sequence
generated by a given discrete memoryless channel (DMC)
\begin{equation}
\label{dmc}
P(y_1,\ldots,y_n|x_1,\ldots,x_n)=\prod_{i=1}^n P(y_i|x_i),~~n=1,2,\ldots.
\end{equation}
When the sequence $\bx$ is sequentially 
fed into a variable--rate finite--state encoder
$\calE=(\calS,f,g)$, this encoder generates 
an infinite sequence of binary strings of variable length,
$\bu=(u_1,u_2,\ldots)$, while going through 
an infinite sequence of states $s_1,s_2,\ldots$
according to
\begin{eqnarray}
u_i&=&f(s_i,x_i),\nonumber\\
s_{i+1}&=&g(s_i,x_i),~~i=1,2,\ldots
\end{eqnarray}
where the initial state
$s_1$ is assumed to be a certain fixed member of $\calS$. 
At the same time and in a similar manner, the finite--state decoder 
$\calD=(\calS',f',g')$
sequentially maps 
$\bu$ and $\by$ to an infinite reproduction 
sequence $\hx_1,\hx_2,\ldots$, $\hx_i\in\hat{\calX}$, 
$i=1,2,\ldots$, with $|\hat{\calX}|=\gamma<\infty$,
using the recursion
\begin{eqnarray}
\hx_{i-d}&=&f'(s_i',u_i,y_i),~~i=d+1,d+2,\ldots\nonumber\\
s_{i+1}'&=&g'(s_i',u_i,y_i),~~i=1,2,\ldots.
\end{eqnarray}
where $d$ (positive integer) is the encoding--decoding 
delay, and the initial state
$s_1'$ is assumed to be a certain fixed member of $\calS'$. 
It is assumed that at each time instant $i$, 
when the decoder is at state $s_i'$, it is able 
to isolate the current input codeword $u_i$ from the following codewords
of the compressed bitstream, 
$u_{i+1},u_{i+2},\ldots$.\footnote{To this end, it would
make sense to assume that the dependence of $u_i$
on $s_i'$ is only via a part of the decoder state
information that is independent of the (random) 
SI sequence, e.g., $s_i''$, which is updated
according to $s_{i+1}''=g''(s_i'',u_i)$.}
To this end, 
we allow a prefix code $\calC(s')$ associated
with each $s'\in\calS'$ (with the option that for some $s'\in \calS'$,
$\calC(s')$ may be empty, in which case, the decoder idles).
For every $u\in\calC(s')$, let $L(u)$ denote
the length of $u$ (in bits). We then assume that the Kraft inequality
\begin{equation}
\sum_{u\in\calC(s')}2^{-L(u)}\le 1
\end{equation}
holds for all $s'\in\calS'$. Note that the above discussion continues to
apply when single codewords are replaced by 
$\ell$--vectors, $u^\ell$, formed by
concatenating $\ell$ (legitimate) codewords successively. In this case,
let $\calC^\ell(s')$ denote the supercode formed by all $\{u^\ell\}$
that originate from state $s'$. Clearly, since the components of $u^\ell$
can be identified recursively, the supercode satisfies the Kraft inequality
as well w.r.t.\ the length function $L(u^\ell)=\sum_{i=1}^\ell L(u_i)$.

For a given single--letter distortion measure $\rho:\calX\times\hat{\calX}\to\reals^+$,
let $\Delta(x^n,\calE,\calD)$ denote the expected distortion 
$\frac{1}{n}\sum_{i=1}^nE\rho(x_i,\hX_i)$
associated with the encoding and decoding $x^n$ by $(\calE,\calD)$,
where the expectation is w.r.t.\ the DMC.
Now, define
\begin{equation}
\Delta_{M,d}(x^n,R)=\min \Delta(x^n,\calE,\calD)
\end{equation}
where the minimum is over all encoder--decoder
pairs $\{(\calE,\calD)\}$ having no more than 
$M$ states each, with delay no longer than $d$, and which satisfy
the rate constraint:
\begin{equation}
\sum_{i=1}^n L(u_i)\le nR.
\end{equation}

While the optimum $(\calE,\calD)$ for 
achieving $\Delta_{M,d}(x^n,R)$ depends, in general,
on $x^n$, we
are interested in a universal algorithm 
(independent of $x^n$) that `competes' with the best
$M$--state encoder--decoder pair $(\calE,\calD)$,
and eventually approaches the {\it operational} (finite--state) W--Z distortion--rate function,
which we define as:
\begin{equation}
\label{opta1}
\Delta(\bx,R)\dfn \lim_{d\to\infty}\lim_{M\to\infty}\limsup_{n\to\infty} \Delta_{M,d}(x^n,R).
\end{equation}
Note that the order of the limits is first over $M$ and then over $d$. This is because
given $M$, the delay $d$ cannot be arbitrarily large. To implement a finite--state machine
with delay $d$, such a system must store the $d$ most recent inputs, which requires the
number of states to be at least exponential in $d$, namely, the maximum possible delay for
a given $M$, which we shall denote by $d_M$, is proportional to $\log M$. A somewhat stronger
notion of the operational finite--state W--Z distortion--rate function is then given by
\begin{equation}
\label{opta2}
\Delta^*(\bx,R)\dfn \lim_{M\to\infty}\limsup_{n\to\infty} \Delta_{M,d_M}(x^n,R).
\end{equation}
Obviously, $\Delta^*(\bx,R)\le \Delta(\bx,R)$.

\section{Long Block Codes are as Good as FSM's}

We start by defining the {\it informational} W--Z distortion--rate function 
(as opposed to the operational definitions of eqs.\ (\ref{opta1}), (\ref{opta2}))
in the following manner:
Let $n$ and $\ell < n$ be two given positive integers,
assume, without essential loss of generality, that $\ell$ divides $n$,
and let us chop the sequence $x^n$ into $\ell$--blocks,
$\{x_{i\ell+1}^{(i+1)\ell}\}_{i=0}^{n/\ell-1}$. Now, let $P(a^\ell)$,
$a^\ell=(a_1,\ldots,a_\ell)\in\calX^\ell$,
denote the empirical probability (relative frequency)
of the $\ell$--vector $a^\ell$
along $x^n$, i.e.,
\begin{equation}
P(a^l)=\frac{\ell}{n}\sum_{k=0}^{n/\ell-1}
1\{x_{i\ell+1}^{(i+1)\ell}=a^\ell\}
\end{equation}
and for any $b^\ell=(b_1,\ldots,b_\ell)\in\calY^\ell$, let
\begin{equation}
P(a^\ell,b^\ell)=
P(a^\ell)P(b^\ell|a^\ell)=
P(a^\ell)\cdot\prod_{i=1}^\ell P(b_i|a_i),
\end{equation}
where $\{P(b_i|a_i)\}$ are the single--letter transition probabilities associated with the
DMC (\ref{dmc}).
Let $(X^\ell,Y^\ell)$ designate random
$\ell$--vectors jointly distributed according to 
$\{P(a^\ell,b^\ell),~a^\ell\in\calX^\ell,~b^\ell\in\calY^\ell\}$,
and define the $\ell$--th order informational 
W--Z distortion--rate function of the source $X^\ell$
w.r.t.\ the SI $Y^\ell$ as follows:
\begin{equation}
\Delta_{X^\ell|Y^\ell}(R)=\min_{U,h} \frac{1}{\ell}E\rho(X^\ell,h(Y^\ell,U))
\end{equation}
where the minimum is over all functions $\{h\}$ from 
$\calY^\ell\times\calU$ to $\hat{\calX}^\ell$, and
over all RV's $\{U\}$ that: (i) take on values 
in an alphabet $\calU$ of size $|\calX|^\ell+1$, (ii)
satisfy the Markov relation $U\to X^\ell\to Y^\ell$, and
(iii) satisfy the inequality
\begin{equation}
H(U)\le \ell R.
\end{equation}

Our main result in this section relates the 
operational W--Z distortion--rate function,
for finite $M$ and $d$, to the $\ell$--th order
informational distortion--rate function associated with the empirical distribution of $x^n$,
as defined above:
\begin{theorem} 
\label{thm1}
For every positive integers $n$ and $\ell$ such that $\ell$ divides $n$:
\begin{equation}
\Delta_{M,d}(x^n,R)\ge 
\Delta_{X^\ell|Y^\ell}\left(R+\frac{2\log M}{\ell}\right)-
\frac{\rho_{\max}d}{\ell},
\end{equation}
where $\rho_{\max}\dfn\max_{x,\hx}\rho(x,\hx)$.
\end{theorem}

Note that the rate--redundancy, $(2\log M)/\ell$, 
depends on the number of states whereas the
distortion redundancy, $\rho_{\max}d/\ell$, depends only the delay. 
However, referring to relationship between $d$ and $M$ (cf.\ the last paragraph of Section 2),
the distortion--redundancy term $\rho_{\max}d/\ell$,
is also bounded by a quantity proportional to $(\log M)/\ell$, like the
rate--redundancy.

Defining now the informational W--Z distortion--rate function of the infinite sequence $\bx$ as
\begin{equation}
\Delta_{\bX|\bY}(R)\dfn\limsup_{\ell\to\infty}\limsup_{n\to\infty}\Delta_{X^\ell|Y^\ell}(R),
\end{equation}
we have the following corollary to Theorem \ref{thm1},
which means that long block codes are asymptotically sufficiently good to attain the
asymptotic performance of general finite--state codes:
\begin{corollary}
\label{cor}
For every $\bx$,
\begin{equation}
\Delta_{\bX|\bY}(R)\ge\Delta^*(\bx,R)\ge 
\Delta_{\bX|\bY}(R^+)\dfn \lim_{R'\downarrow R}\Delta_{\bX|\bY}(R').
\end{equation}
\end{corollary}
Observe that, by monotonicity, 
$\Delta_{\bX|\bY}(R^+)=\Delta_{\bX|\bY}(R)$ 
(hence the two inequalities become equalities) for every $R\ge 0$,
with the possible exception of a countable set of points.
Corollary \ref{cor} follows from Theorem 1 
in a simple manner: For any given $R'> R$,
Theorem 1 implies that 
$\Delta_{M,d_M}(x^n,R)\ge\Delta_{X^\ell|Y^\ell}(R')-\rho_{\max}d_M/\ell$ 
for all sufficiently large $\ell$.
Taking first, the limsup as $n\to\infty$, next the limsup as $\ell\to\infty$, then 
the limit as $M\to \infty$, and finally, the limit $R'\to R$, gives the right inequality.
The left inequality follows from the fact that a block code of length $\ell$ is implementable
by an FSM with $\alpha^\ell$ states (cf.\ \cite[Example 2, p.\ 406]{Ziv78}) thus
$\Delta_{X^\ell|Y^\ell}(R)\ge\Delta_{\alpha^\ell,d_{\alpha^\ell}}(x^n,R)$, 
and the result is obtained
by taking first the limsup over $n$ and then the limsup over $\ell$ at both sides of this
inequality. 

We now turn to the proof of Theorem \ref{thm1}.
\newline\noindent
{\it Proof.} For a given combination of an $M$--state
encoder and an $M$--state decoder $(\calE,\calD)$, consider
the joint probability distribution
\begin{equation}
P(a^\ell,c^\ell,s,s')=\frac{\ell}{n}\sum_{i=0}^{n/\ell-1}
1\{x_{i\ell+1}^{i\ell+\ell}=a^\ell,
u_{i\ell+1}^{i\ell+\ell}=c^\ell,
s_{i\ell+1}=s,s_{i\ell+1}'=s'\},
\end{equation}
for every
$a^\ell\in\calX^\ell$, $s\in\calS$, $s'\in\calS'$, and $c^\ell\in\calC^\ell(s')$.
Note that $\hx_{i\ell-d+1}^{i\ell-d+\ell}$ depends (deterministically) only on
$y_{i\ell+1}^{i\ell+\ell}$, $u_{i\ell+1}^{i\ell+\ell}$, and $s_{i\ell+1}'$.
Let us denote then 
$\hx_{i\ell-d+1}^{i\ell-d+\ell}=
h(y_{i\ell+1}^{i\ell+\ell},u_{i\ell+1}^{i\ell+\ell},s_{i\ell+1}')$,
and assuming that $\ell > d$, 
let $\hx_{i\ell+1}^{i\ell-d+\ell}\dfn 
h'(y_{i\ell+1}^{i\ell+\ell},u_{i\ell+1}^{i\ell+\ell},s_{i\ell+1}')$
be defined simply by truncating the first $d$ components of 
$h(y_{i\ell+1}^{i\ell+\ell},u_{i\ell+1}^{i\ell+\ell},s_{i\ell+1}')$.
Now, define
\begin{eqnarray}
P(a^\ell,b^\ell,c^\ell,\hat{a}^{\ell-d},s,s')&=&
P(a^\ell,c^\ell,s,s')P(b^\ell|a^\ell)
1\{\hat{a}^{\ell-d}=h'(b^\ell,c^\ell,s')\}\nonumber\\
&=&P(a^\ell,c^\ell,s,s')[\prod_{i=1}^\ell P(b_i|a_i)]
1\{\hat{a}^{\ell-d}=h'(b^\ell,c^\ell,s')\}, 
\end{eqnarray}
for every $b^\ell\in\calY^\ell$, and $\hat{a}^{\ell-d}\in\hat{\calX}^{\ell-d}$.
Let $(X^\ell,Y^\ell,U^\ell,\hX^{\ell-d},S,S')$ designate a random
vector that is distributed according to this joint probability mass function.
Now, accoding to the rate constraint:
\begin{eqnarray}
\label{17}
nR&\ge&\sum_{i=1}^n L(u_i)\nonumber\\
&=&\sum_{i=0}^{n/\ell-1} L(u_{i\ell+1}^{i\ell+\ell})\nonumber\\
&=&\frac{n}{\ell}\sum_{s'\in\calS'}\sum_{c^\ell\in\calC^\ell(s')} 
P(c^\ell,s')L(c^\ell)\nonumber\\
&\ge&\frac{n}{\ell}\sum_{s'\in\calS'}\sum_{c^\ell\in\calC^\ell(s')} 
P(c^\ell,s')\log\frac{1}{P(c^\ell|s')}
\end{eqnarray}
where the last step follows from the postulate that 
$\{L(c^\ell),~c^\ell\in\calC^\ell(s')\}$ satisfy the Kraft inequality
for each $s'$. It follows then that
$H(U^\ell|S')\le \ell R$. Therefore,
\begin{eqnarray}
\label{18}
\ell R &\ge&H(U^\ell|S')\nonumber\\
&=&H(U^\ell,S')-I(S';U^\ell)\nonumber\\
&\ge&H(U^\ell,S')-H(S')\nonumber\\
&\ge&H(U^\ell,S')-\log M.
\end{eqnarray}
Next, observe that 
$u_{i\ell+1}^{i\ell+\ell}$ depends solely on $s_{i\ell+1}$ and
$x_{i\ell+1}^{i\ell+\ell}$ but not on $y_{i\ell+1}^{i\ell+\ell}$, and so, 
$U^\ell\to (X^\ell,S)\to Y^\ell$ is a Markov chain.
But since $S\to X^\ell\to Y^\ell$ is also a 
Markov chain (due to the DMC), then so is
$(U^\ell,S)\to X^\ell\to Y^\ell$. In addition, $\hX^{\ell-d}=h'(Y^\ell,U^\ell,S')$.
It therefore follows that
\begin{eqnarray}
\frac{1}{n}E\rho(x^n,\hat{X}^n)
&\ge&\frac{1}{\ell}E\rho(X^{\ell-d},\hat{X}^{\ell-d})\nonumber\\
&\ge&\frac{1}{\ell}[E\rho(X^\ell,\hat{X}^\ell)-d\cdot\rho_{\max}]
\end{eqnarray}
where $\hX^\ell$ is defined by concatenating $\hX^{\ell-d}$ with a
random $d$--vector in $\hat{\calX}^d$ 
that is an arbitrary function of $(Y^\ell,U^\ell,S')$.
Next, observe that
\begin{eqnarray}
\label{20}
\ell R&\ge& H(U^\ell,S')-\log M\nonumber\\
&\ge&H(U^\ell,S,S')-2\log M,
\end{eqnarray}
that is,
\begin{equation}
H(U^\ell,S,S')\le\ell\left(R+\frac{2\log M}{\ell}\right),
\end{equation}
and the Markovity of the chain $(U^\ell,S)\to X^\ell\to Y^\ell$ implies
Markovity of $(U^\ell,S,S')\to X^\ell\to Y^\ell$.
Moreover, for the purpose of deriving 
a lower bound, let us also allow $h$ to depend on $S$
too, i.e., $\hat{X}^\ell=h(Y^\ell,U^\ell,S,S')$. 
We now have the following lower bound to $\Delta_{M,d}(x^n,R)$:
\begin{equation}
\Delta_{M,d}(x^n,R)\ge \min_{U^\ell,S,S'h} 
\frac{1}{\ell} E\rho(X^\ell,h(Y^\ell,U^\ell,S,S'))-\frac{d\rho_{\max}}{\ell}
\end{equation}
where the minimum is subject to the constraints:
\begin{equation}
H(U^\ell,S,S')\le\ell\left(R+\frac{2\log M}{\ell}\right)
\end{equation}
and
\begin{equation}
(U^\ell,S,S')\to X^\ell\to Y^\ell~\mbox{is a Markov chain.}
\end{equation}
Now, observe that in this minimization problem 
$U^\ell$, $S$, and $S'$ appear always
together. Let us define then $U\dfn (U^\ell,S,S')$ and 
further reduce this expression by taking the minimum of
the distortion over all $(U,h)$ subject 
to the constraints 
$H(U)\le \ell(R+2\log M/\ell)$, 
and $U\to X^\ell\to Y^\ell$,
which is $\Delta_{X^\ell|Y^\ell}(R+2\log M/\ell)$
by definition.
This completes the proof of the Theorem \ref{thm1}.
$\Box$

We now describe a (universal) block coding 
scheme that asymptotically (for large $\ell$) achieves
$\Delta_{\bX|\bY}(R)$ and hence also 
$\Delta^*(\bx,R)$ (for almost all values of $R$):
Given $x^n$, compute its $\ell$--th order empirical
distribution, $\{P(a^\ell),~a^\ell\in\calX^\ell\}$ (which is $P(X^\ell)$), and
find the RV $U$ and the function $h$ that achieve $\Delta_{X^\ell|Y^\ell}(R)$. 
The (stochastic) encoder applies the channel 
$P(U|X^\ell)$ to every $\ell$--vector 
$x_{i\ell+1}^{i\ell+\ell}$ and then performs
entropy coding according to the marginal of $U$, 
after transmitting a header that describes
the entropy coding rule of the optimum $U$ 
(which depends only on the marginal of $U$)
and the function $h$, which together require no more than
$\lceil \log(n/\ell+1)^{\alpha^\ell}\rceil$ bits, 
which is log of the number of different
empirical distributions of
superletters formed by $\ell$--vectors). The decoder first decodes the header, 
then $U$, and
finally, reconstructs the source by applying $\hat{X}^\ell=h(Y^\ell,U)$. 
The rate is upper bounded by
\begin{equation}
\frac{1}{n}\{\lceil\log[(k+1)^{\alpha^\ell}]\rceil+
\frac{n}{\ell}[H(U)+1]\}\nonumber\\
\le R+\frac{1}{\ell}+\frac{\alpha^\ell}{n}\log\left(\frac{n}{\ell}+1\right)
\end{equation}
where we have used the fact that $H(U)\le \ell R$. Clearly, if $\ell$ is
large and $n >> \alpha^\ell$, this is arbitrarily close to $R$. The distortion
$\Delta_{X^\ell|Y^\ell}(R)$ is maintained by definition.

Note that complexity of this scheme is mostly in 
the optimization over $h$ and $U$,
which is not negligible, but is a 
function of $\ell$ only. This is in contrast to the 
schemes proposed in \cite{Ziv80},\cite{Ziv84}, which require an 
exhaustive search over sets of sequences of length
$n (>> \alpha^\ell)$, namely, complexity that 
grows exponentially with $n$. The stochastic
encoder $X^\ell\to U$ 
can also be implemented deterministically, but then the encoding 
complexity will be exponential in $n$:
Select independently at random a set of $M=2^{(n/\ell)[I(X^\ell;U)+\epsilon]}$
vectors $U^n(i)$, $i=1,\ldots,M$. Given $x^n$, 
find a jointly typical vector $U^n(i)$
(in the superalphabet of $\ell$--vectors),
and transmit it using 
$(n/\ell)[I(X^\ell;U)+\epsilon]\le (n/\ell)[H(U)+\epsilon]$ bits
plus a (relatively small) header that describes the 
type class of $x^n$, from which the decoder 
can also figure
out $h$ on its own.
By the Markov lemma, with high probability,
$(U,X^\ell,Y^\ell)$ will be jointly typical 
and hence $\hat{X}^n$ will satisfy 
the distortion constraint.

\noindent
{\bf Discussion.} Three comments are in order:
\begin{itemize}
\item [1.]
When $\bx$ emerges from a discrete memoryless source (DMS), rather than 
being an individual sequence, it is well known that the distortion--rate
function is the classical W--Z distortion rate function for that DMS,
$\Delta_{X|Y}^{WZ}(R)$. Clearly, 
by analyzing the performance of block codes (rather than FSM's) on the given DMS,
using the same tecnique as above, one can show that
\begin{equation}
\Delta_{X|Y}^{WZ}(R)=
\inf_{\ell\ge 1}\Delta_{X^\ell|Y^\ell}(R),
\end{equation}
where now $X^\ell$ designates a random $\ell$--vector from the source.
This is true because for both sides of this equality, we have a direct
theorem and a converse theorem.
\item [2.]
In view of item no.\ 1 above and our direct thoerem, we have actually shown
that it is possible to approach the W--Z rate--distortion function 
(of both a DMS and an individual sequence) without binning.
\item [3.]
The above result extends to a model of scalable coding (successive
refinement), in analogy to \cite{SM04}. The details and the discussion appear in the Appendix.
\end{itemize}

\section{Universalilty and the Critical Growth Rate of $M$}

In the previous section, we have considered
a reference class of encoders and decoders 
with a {\it fixed} number of states, $M$, and a {\it fixed} delay, $d$, and
we have shown that the (operative) rate--distortion 
function w.r.t.\ this class can be approached by
using (sufficiently long) block codes. 
In this section, we address a somewhat different question,
pertaining to a regime that allows both the 
number of states and the delay to {\it grow} with 
$n$, the length of the input sequence $x^n$. That is, $M=M_n$ and $d=d_n$.
For the sake of simplicity, we will be `generous' with
regard to the delay, allowing it to be maximum, namely, $d_n=d_{M_n}$, 
and then we focus on the following question: What is the
highest growth rate of $M_n$ as a function of $n$, below which
it is still possible to universally attain (in a 
sense to be defined soon), using any general block code for $n$--sequences, 
the performance of the best
encoder--decoder with $M_n$ states and delay $d_{M_n}$?

For the sake of convenience, in this section, 
instead of confining ourselves to either rate--distortion
functions or distortion--rate functions, we will treat 
rate and distortion in a more symmetric fashion
by defining achievable pairs $(R,\Delta)$ in the spirit of
the definitions in Section 1: 
Given $x^n$, a pair $(R,\Delta)$ is said to be $M_n$--{\it achievable}
if there exists an $M_n$--state encoder $\calE=(\calS,f,g)$ and a 
$M_n$--state decoder $\calD=(\calS',f',g')$, with overall
delay not exceeding $d_n=d_{M_n}$, such that $\sum_{i=1}^n L(u_i)\le nR$ 
and $\sum_{i=1}^n E\rho(x_i,\hat{X}_i)\le n\Delta$.
Referring to the previous definitions, given $x^n$, the pair 
$(R,\Delta_{M_n,d_{M_n}}(x^n,R))$ is always $M_n$--achievable.

While the definition of an achievable pair $(R,\Delta)$ allows the 
choice of an encoder--decoder that depends on $x^n$, 
we now define the notion of a {\it universally} achievable
pair $(R,\Delta)$: A pair $(R,\Delta)$ is said to 
be {\it universally achievable} w.r.t.\ $\{M_n\}_{n\ge 1}$
if for every $\epsilon > 0$, $\delta > 0$, 
and $n$ sufficiently large, there exists
an encoder--decoder that achieves rate less than 
or equal to $R+\epsilon$ and distortion less than
or equal to $\Delta+\delta$ for every $x^n$ 
for which $(R,\Delta)$ is $M_n$--achievable.

Let us assume, from now on, that $M_n$ 
grows asymptotically linearly with some power of $n$,
that is, $\lim_{n\to\infty} 
(\log M_n)/\log n = \theta$, where $\theta$ is a certain
positive real. In this section, we are interested 
in the critical value of $\theta$ below which
universal achievability of {\it every} pair 
$(R,\Delta)$ is guaranteed, but above this critical
value, there exist pairs $(R,\Delta)$ that are not universally achievable.

The following two theorems tell us that this 
critical value is $\theta=1$, in other words, the
critical asymptotic growth rate of $M_n$ is {\it linear}.

\begin{theorem} (Direct Theorem): If $\theta < 1$, 
then every $(R,\Delta)$ is universally
achievable.
\end{theorem}

\noindent
{\it Proof.} Assume, for the sake of simplicty and
without essential loss in generality, that $M_n=n^\theta$, $\theta < 1$, and
consider the following mechanism for an encoding $x^n$: 
The encoder examines all possible pairs $\{(\calE,\calD)\}$
with $M_n$ states, on the given $x^n$, 
and computes the coding rate and the expected
distortion (w.r.t.\ the randomness of the 
known channel from $x^n$ to $y^n$). If it
finds an encoder--decoder pair $(\calE^*,\calD^*)$ that achieves a 
specified rate--distortion pair $(R,\Delta)$, it first transmits a header
with the description of $\calD^*$ and then 
encodes $x^n$ using $\calE^*$ (if it does
find such an encoder then $(R,\Delta)$ 
is not $M_n$--achievable in the first place).
The decoder, after decoding the index of the 
decoder $\calD^*$, uses this decoder to
produce the reproduction $\hat{x}^n$ based 
on $y^n$ and the remaining part of the bitstream.
Obviously, the distortion associated with such an 
encoder is the same as that of 
$(\calE^*,\calD^*)$, namely, less than 
or equal to $\Delta$. The rate is the same as that of
$(\calE^*,\calD^*)$ (which means less than or 
equal to $R$) plus the rate associated with
the header. Thus, it remains to show that 
the normalized redundancy associated with the header
goes to zero as $n\to\infty$ whenever 
$\theta < 1$. To this end, we now evaluate the number of
bits necessary to describe the decoder with $\calD^*$. 

First, observe that for each $s_i'\in\calS'$, 
$u_i$ may take values in a prefix code $\calC(s_i')$,
which can be described by a tree. As each such tree contains
at most $\alpha$ leaves, and there 
are $(k-1)!$ different trees with $k$ leaves, the total
number of possible trees is 
$K=\sum_{k=1}^\alpha (k-1)!$, thus the description of each such
tree takes $\lceil \log K \rceil$ bits, and since 
such a tree should be specified for every
$s'\in\calS'$ this takes $M_n\cdot\lceil \log K \rceil$ 
bits altogether. Next, the function
$f'$ should be described. As there are no more 
than $\gamma^{M_n\alpha\beta}$ functions $\{f'\}$
for every possible tree, the description 
of $f'$ takes $\lceil M_n\alpha\beta\log\gamma\rceil$ bits.
Similarly, the description of $g'$ requires 
at most $\lceil M_n\alpha\beta\log M_n\rceil$ bits.
Thus, the total number of bits associated with the header 
is then $O(n^\theta\log n)$, which when normalized by $n$ (to get
the redundancy per source symbol), 
tends to zero as $n\to \infty$, since $\theta$ is assumed
strictly less than unity. 
This completes the proof of the Theorem.
$\Box$

For the converse part, we will make 
the additional assumptions that $\hat{\calX}=\calX$,
$\calX$ is a group with an addition operation 
(modulo $\alpha$), and that the distortion
function $\rho(x,\hx)$ depends on $x$ and $\hx$ 
only via their difference $x-\hx$ 
(w.r.t.\ the group arithmetic). 
We will then denote $\rho_0(x-\hx)=\rho(x,\hx)$, 
where $\rho_0:\calX\to\reals^+$.

\begin{theorem} 
\label{conv}
(Converse Theorem): Under 
the assumptions of the last paragraph,
if $\theta > 1$ there exist pairs
$(R,\Delta)$ that are not universally
achievable.
\end{theorem}

\noindent
{\bf Discussion.} An alternative question with regard to
universality w.r.t.\ FSM's with a growing number of states, 
which is closer in spirit to the results of Section 3,
could have been the following: What is
the critical growth rate of $M_n$ that still allows 
$\Delta_{M_n,d_{M_n}}(x^n,R)$ to be achievable by a universal
code for all $x^n$? This definition is seemingly stronger because
it appears to give rise to `adaptation' of the distortion to the given $x^n$
rather than making a commitment to a fixed distortion level $\Delta$, regardless
of $x^n$. However, it is easy to see that both our direct theorem and
converse theorem are suitable for this definition too, and therefore,
so is the conclusion regarding the linear critical rate of $M_n$.
The direct part would be the same, but with $(\calE^*,\calD^*)$ being
defined as the encoder--decoder pair that achieves
$\Delta_{M_n,d_{M_n}}(x^n,R)$, or alternatively,
$R_{M_n,d_{M_n}}(x^n,\Delta)$, the $M_n$--state rate--distortion function,
defined in a dual manner.
Regarding the converse, as we demonstrate
in the proof of Theorem \ref{conv} below, for every given $\theta> 1$,
there is a pair $(R,\Delta)$
which is $M_n$--achievable for certain sequences, and hence
$\Delta_{M_n,d_{M_n}}(x^n,R)\le \Delta$, or, equivalently,
$R_{M_n,d_{M_n}}(x^n,\Delta)\le R$. But no single encoder performs
even close to 
$R_{M_n,d_{M_n}}(x^n,\Delta)$ simultaneously for all these sequences.

\vspace{0.5cm}

\noindent
{\it Proof of Theorem \ref{conv}.} Assume, again, that $M_n=n^\theta$. 
We will now show that for $\theta > 1$, there
exists a rate--distortion pair $(R,\Delta)$ 
which is not universally achievable. For a given
$\Delta$, let 
\begin{equation}
\phi(\Delta)=\max\{H(Z):~E\rho_0(Z)\le\Delta\},
\end{equation}
where $H(Z)$ is the entropy of an RV $Z$ taking on values in $\calX$.
Let $R_{X|Y}^{WZ}(\Delta)$ denote the W--Z rate--distortion
function of the memoryless uniform source $X$ with side information
$Y$ (generated by the given DMC $P(y|x)$) w.r.t.\ $\rho$. For a given
$\Delta$, and a
given $\theta > 1$, let 
\begin{equation} 
R=\frac{\phi(\Delta)}{\theta-1}
\end{equation}
and select $\Delta$
to be sufficiently small such that
\begin{equation}
R=\frac{\phi(\Delta)}{\theta-1} < 
R_{X|Y}^{WZ}(\Delta)\le R_X(\Delta)=\log\alpha-\phi(\Delta),
\end{equation}
where $R_X(\Delta)$ is the ordinary rate--distortion function (without side information) of the
memoryless uniform source $X$ w.r.t.\ $\rho$.
We wish to show that for such a choice of $R$ and $\Delta$,
there exists a set of sequences $\{x^n\}$ 
for each of which $(R,\Delta)$ is $M_n$--achievable, but
on the other hand, there is no {\it single} code 
that simultaneously achieves $(R,\Delta)$ for all
sequences in this set.

For the above choice of $R$ and $\Delta$, 
consider a random process defined as follows.
Let $m$ be the
the solution to the equation
\begin{equation}
2^{Rm} =\frac{n}{m},
\end{equation}
and assume that this solution is integer. Further, let $\calF$ be a 
set of $2^{mR}$ $m$--vectors $\calF=\{\bu_1 ,\ldots,\bu_{2^{mR}}\}$,
$\bu_i,\in\calX^m$, $i=1,\ldots,2^{mR}$. 
Assuming that $m$ divides $n$ (i.e., $2^{mR}$ is integer), let $x^n$ be formed
by concatenatingg $n/m$ $m$--vectors $\{x^m(i),~i=1,\ldots,n/m\}$, where 
\begin{equation}
\label{process}
x^m(i)=u^m(i)+z^m(i),~~~i=1,\ldots,n/m,
\end{equation}
$u^m(i)$ is an arbitrary member of $\calF$,
and $z^m(i)\in\calX^m$ is a vector of
i.i.d.\ random variables, each component of which is 
distributed according to the distribution
$P^*$ on $\calX$ which achieves $\phi(\Delta)$. 
Now, we further assume that $\calF$ is a 
good\footnote{In the sense of small error probability.} code
for the additive memoryless 
channel $X=U+Z$, where $U$ designates the channel input at rate $R$, 
$Z\sim P^*$ is the memoryless noise, and $X$ 
stands for the channel output. Since $R$ is assumed less than the capacity of
this channel, given by $C=R_X(\Delta)=\log\alpha-\phi(D)$,
then such a good code exists. This means that upon observing
$x^m(i)$, one can identify $u^m(i)$ correctly with high probability, 
provided that $m$ is large. 

Consider next a {\it block} code of length $m$ operating on $x^n$ as follows: 
For every $i=1,\ldots,n/m$,
the encoder first decodes $u^m(i)$ from $x^m(i)$, 
and then transmits a description of the
decoded version, say $\hu^m(i)$,
using $\log|\calF|=mR$ bits. The decoder, in
turn, reconstructs $\hat{x}^m(i)=\hu^m(i)$ 
(without using the side information). Since
$\hu^m(i)=u^m(i)$ with high probability, 
then the distortion between $x^m(i)$ and $\hx^m(i)$,
is about $\Delta$, because the noise 
$z^m(i)$ is distributed according to $P^*$, whose
$\rho_0$--moment does not exceed $\Delta$.
This means that the pair $(R,\Delta)$ is essentially achievable by
a (non--universal) block code of size $m$.

Now, since the set of typical input $m$--vectors to the
block code is of size that is of 
the exponential order of $2^{m[R+\phi(\Delta)]}$,
this block code can be implemented by an FSM with with essentially no 
more than $m2^{m[R+\phi(\Delta)]}$ states. This can be done by
constructing an (incomplete) $\alpha$--ary
context--tree with $2^{m[R+\phi(\Delta)]}$ leaves, corresponding
to the various typical sequences, where 
the state set is the set of nodes plus the leaves
of this incomplete tree.\footnote{To be precise, one should add $m$ more
states corresponding to a modulo--$m$ time counter, in order to idle
for non-typical sequences (which are not terminated by the leaves)
until the end of the block and then submit
an error message.}
Thus, the number of states 
as a function of $n$ is given by
\begin{eqnarray}
M_n &\le& m2^{m[R+\phi(\Delta)]}\nonumber\\
&\le& [m2^{mR}]^{[1+\phi(\Delta)/R]}\nonumber\\
&=&n^{1+\phi(\Delta)/R}\nonumber\\
&=&n^\theta.
\end{eqnarray}
This means that we have shown that for 
the above choice of $R$ and $\Delta$, the pair
$(R,\Delta)$ is $n^\theta$--achievable 
for every $x^n$ that is typical to the above 
defined process.

We now argue that no block--code of length $n$ can attain $(R,\Delta)$
simulatenously for {\it all} typical 
sequences of the process (\ref{process}), and for
all `good' channel codes $\{\calF\}$, which
yield error probability less than 
$2^{-m[E_r(R)-\delta]}$, where $E_r(R)$ is the random
coding error exponent \cite{Gallager68}
of the channel $X=U+Y$ w.r.t.\ the uniform random coding
input distribution, and $\delta\in(0,E_r(R))$ 
is arbitrary.\footnote{Such a code may
harm the rate (beyond $R$) and the distortion (beyond $\Delta$)
only for fraction $2^{-m[E_r(R)-\delta]}$ of the $m$--segments.}
Furthermore, we show that even $(R_{X|Y}^{WZ}(\Delta)-\epsilon,\Delta)$,
for arbitrarily small (but fixed) 
$\epsilon > 0$, cannot be simultaneously attained
for all those sequences (recall that $R_{X|Y}^{WZ}(\Delta) > R$).

Let us denote the set of all the typical 
$u$--sequences by $\calG$, i.e., $\calG$ is the 
set of all sequences $\{u^n\}$ whose segments $\{u^m(i)\}$ form a code 
(for the channel $X=U+Y$) whose error probability, $P_e(u^n)$,
is below $2^{-m[E_r(R)-\delta]}$, and observe that
\begin{equation}
\mbox{Pr}\{\calG^c\}=\mbox{Pr}\{u^n:~P_e(u^n) > 2^{-m[E_r(R)-\delta]}\}
\le\frac{E\{P_e(U^n)\}}{2^{-m[E_r(R)-\delta]}}\le 
\frac{2^{-mE_r(R)}\}}{2^{-m[E_r(R)-\delta]}}
=2^{-m\delta},
\end{equation}
where the first inequality follows from the Chebychev inequality.
Now, assume conversely, that there exists a source code that {\it does} 
achieve $(R_{X|Y}^{WZ}(\Delta)-\epsilon,\Delta)$ 
for all $x^n$ induced by all $u^n\in\calG$ and all
typical $z^n$--sequences, namely, sequences for which (a very high fraction
of) the $m$-segments $\{z^m(i)\}$ are $P^*$--typical.
Then, we have
\begin{equation}
\frac{1}{|\calG|}\sum_{u^n\in \calG} EL(u^n+Z^n) 
\le n[R_{X|Y}^{WZ}(\Delta)-\epsilon]
\end{equation}
and
\begin{equation}
\frac{1}{|\calG|}\sum_{u^n\in \calG} E\rho(u^n+Z^n,\hat{X}^n) \le n\Delta,
\end{equation}
and so, for every $\lambda \ge 0$,
\begin{equation}
\frac{1}{|\calG|}\sum_{u^n\in \calG} [EL(u^n+Z^n)+
\lambda E\rho(u^n+Z^n,\hat{X}^n)]
\le n[R_{X|Y}^{WZ}(\Delta)-\epsilon+\lambda\Delta],
\end{equation}
where the inner expectations are w.r.t.\ the uniform distribution over
all typical $z$--sequences.
Consider now a {\it random} selection of 
the $2^{Rm}=n/m$ members of $\calF$
independently and with uniform distribution over $\calX^m$.
On the one hand, this induces the uniform distribution over $\calX^n$,
for which we know, by the converse to the W--Z
rate--distortion theorem, that either
\begin{equation}
\frac{1}{\alpha^n}\sum_{u^n} E\rho(u^n+Z^n,\hat{X}^n) \ge n\Delta,
\end{equation}
or
\begin{equation}
\frac{1}{\alpha^n}\sum_{u^n} EL(u^n+Z^n) \ge nR_{X|Y}^{WZ}(\Delta),
\end{equation}
where the inner expectations over $Z^n$ are as before (note that 
as $U^n$ is uniformly distributed then so is $X^n=U^n+Z^n$ regardless
of $Z^n$, which is independent).
It then follows that there exists 
$\lambda \ge 0$ (which is bounded independently of $n$) such that
\begin{equation}
\label{lag}
\frac{1}{\alpha^{n}}\sum_{u^n} 
[EL(u^n+Z^n)+
\lambda E\rho(u^n+Z^n,\hat{X}^n)]
\ge n[R_{X|Y}^{WZ}(\Delta)+\lambda\Delta].
\end{equation}
To see why this is true, let us denote
$n\Delta'\dfn\alpha^{-n}\sum_{u^n}E\rho(u^n+Z^n,\hat{X}^n)$, 
and then the left--hand side of eq.\ (\ref{lag}) is lower bounded by
$n[R_{X|Y}^{WZ}(\Delta')+\lambda\Delta']$. Now, if $\Delta'\le \Delta$, then
eq.\ (\ref{lag}) clearly holds for $\lambda=0$. Else, if $\Delta' > \Delta$, it obviously holds for
\begin{eqnarray}
\lambda&=&\frac{R_{X|Y}^{WZ}(\Delta)-R_{X|Y}^{WZ}(\Delta')}{\Delta'-\Delta}\nonumber\\
&\le&\frac{R_{X|Y}^{WZ}(0)-R_{X|Y}^{WZ}(\Delta)}{\Delta-0}\nonumber\\
&=& \frac{H(X|Y)-R_{X|Y}^{WZ}(\Delta)}{\Delta}
\end{eqnarray}
where the inequality follows from the convexity of
the Wyner--Ziv rate--distortion function 
\cite{WZ76}, \cite[p.\ 439, Lemma 14.9.1]{CT91}. Therefore, in either case, the 
value of $\lambda$ that satisfies eq.\ (\ref{lag}) is bounded independently
of $n$ (for $\Delta > 0$).
For this value of $\lambda$, we then have
\begin{eqnarray}
\label{lower}
&& \sum_{u^n\in \calG^c}
[EL(u^n+Z^n)+
\lambda E\rho(u^n+Z^n,\hat{X}^n)]\nonumber\\
&=& \sum_{u^n}
[EL(u^n+Z^n)+
\lambda E\rho(u^n+Z^n,\hat{X}^n)]-\nonumber\\
& &\sum_{u^n\in \calG}
[EL(u^n+Z^n)+
\lambda E\rho(u^n+Z^n,\hat{X}^n)]\nonumber\\
&\ge&n\alpha^{n}[R_{X|Y}^{WZ}(\Delta)+\lambda\Delta]-
n|\calG|[R_{X|Y}^{WZ}(\Delta)-\epsilon+\lambda\Delta]\nonumber\\
&\ge&n\alpha^{n}[R_{X|Y}^{WZ}(\Delta)+\lambda\Delta]-
n\alpha^{n}[R_{X|Y}^{WZ}(\Delta)-\epsilon+\lambda\Delta]\nonumber\\
&=&n\epsilon\alpha^{n}.
\end{eqnarray}
On the other hand, assuming that $\max_x\rho_0(x)=\rho_{\max} < \infty$,
and that $\max_{x^n}L(x^n)\le nL$ for some finite $L > 0$ (otherwise, long
codewords would be better
transmitted uncompressed plus an extra bit to tell that
they are uncompressed), then we have
\begin{eqnarray}
\label{upper}
\sum_{u^n\in \calG^c}
[EL(u^n+Z^n)+
\lambda E\rho(u^n+Z^n,\hat{X}^n)]
&\le&n|\calG^c|(L+\lambda\rho_{\max})\nonumber\\
&=&\mbox{Pr}\{\calG^c\}n\alpha^{n}(L+\lambda\rho_{\max})\nonumber\\
&\le&n2^{-m\delta}\alpha^{n}(L+\lambda\rho_{\max}).
\end{eqnarray}
Comparing now the right--most sides 
of eqs.\ (\ref{lower}) and (\ref{upper}), we get
$\epsilon \le (L+\lambda\rho_{\max})2^{-m\delta}$, 
which is a contradiction for all large
$n$ (and $m$) since $\epsilon > 0$ was assumed a
constant (and $\lambda$ is bounded). Thus, we have disproved
the existence of a code that achieves 
$(R,\Delta)$ simulateneously for all typical sequences
of the process described above. This completes the proof.

\section*{Appendix}
\renewcommand{\theequation}{A.\arabic{equation}}
    \setcounter{equation}{0}

\begin{center}
{\bf Extending the Results of Section 3 to Successive Refinement Codes}
\end{center}

Consider a two--stage coding scheme with a successive refinement structure.
The first stage is as in Section 3. In the second stage, 
there is an additional finite--state encoder $\calE'$ 
that transmits a refining description $\bv=(v_1,v_2,\ldots)$
of variable--length binary strings (similarly as $\bu$),
and an additional finite--state decoder $\calD'$ that
has access to both $\bu$ and $\bv$ as 
well as to another SI sequence $\bz=(z_1,z_2,\ldots)$.
It is assumed that 
\begin{equation}
P(y^n,z^n|x^n)=\prod_{i=1}^nP(y_i,z_i|x_i),~~~ n=1,2,\ldots
\end{equation}
More precisely, the description of the second stage is as follows:
When the sequence $\bx$ is sequentially
fed into a variable--rate finite--state encoder
$\calE'=(\calT,p,g)$, this encoder generates
an infinite sequence of binary strings of variable length,
$\bv=(v_1,v_2,\ldots)$, while going through
an infinite sequence of states $t_1,t_2,\ldots$
according to
\begin{eqnarray}
v_i&=&p(t_i,x_i),\nonumber\\
t_{i+1}&=&q(t_i,x_i),~~i=1,2,\ldots
\end{eqnarray}
where the initial state
$t_1$ is assumed to be a certain fixed member of $\calT$.
At the same time and in a similar manner, 
the second--stage finite--state decoder
$\calD'=(\calT',p',q')$
sequentially maps
$\bu$, $\bv$ and $\bz$ to an infinite reproduction
sequence $\tx_1,\tx_2,\ldots$, using
the recursion
\begin{eqnarray}
\tx_{i-d'}&=&p'(t_i',u_i,v_i,z_i),~~i=d'+1,d'+2,\ldots\nonumber\\
t_{i+1}'&=&q'(t_i',u_i,v_i,z_i),~~i=1,2,\ldots.
\end{eqnarray}
where $d'$ (positive integer) is the 
second--stage encoding--decoding delay, and the initial state
$t_1'$ is assumed to be a certain fixed member of $\calT'$.
Similarly to the model of variable--rate coding of the first stage, 
it is assumed that at each time instant $i$,
when state decoder is at state $t_i'$ 
and reads\footnote{The second stage decoder can keep
a copy of $s_i''$ of footnote no.\ 2 as it has access to $\{u_i\}$ as well.}
the current first--stage codeword $u_i$, it is able
to isolate the current input codeword $v_i$ from the following codewords
of the compressed bitstream, $v_{i+1},v_{i+2},\ldots$. To this end,
we allow a prefix code $\calC'(t',u)$ associated
with each 
$t'\in\calT'$ and $u\in\calC(s')$. 
For every $v\in\calC'(t')$, let $L'(v)$ denote
the length of $v$ (in bits). We then assume that the Kraft inequality
\begin{equation}
\sum_{v\in\calC'(t',u)}2^{-L'(v)}\le 1
\end{equation}
holds for all $t'\in\calT'$, $u\in\calC(s')$. 
Again, this discussion continues to
apply when single codewords are replaced by
$\ell$--vectors, $v^\ell$, formed by
concatenating $\ell$ (legitimate) codewords successively. In this case,
let $[\calC']^\ell(t',u^l)$ denote the supercode formed by all $\{v^\ell\}$
that originate from $(t',u^l)$. Clearly, since the components of $v^\ell$
can be identified recursively, the supercode satisfies the Kraft inequality
as well w.r.t.\ the length function $L'(v^\ell)=\sum_{i=1}^\ell L'(v_i)$.

Let $\Delta'(x^n,\calE,\calD)$ denote the expected distortion
$\frac{1}{n}E\rho'(x^n,\tilde{X}^n)
\dfn\frac{1}{n}\sum_{i=1}^nE\rho'(x_i,\tilde{X}_i)$
associated with the second--stage encoding and decoding of $x^n$ by 
$(\calE,\calD)$ and
$(\calE',\calD')$,
where the expectation is w.r.t.\ the DMC's. 
For a given $x^n$ and rate pair $(R,\Delta R)$,
a distortion pair $(D_1,D_2)$ is said to 
be $(M,d,d')$--achievable if there exist
encoder--decoders
$(\calE,\calD)$ and
$(\calE',\calD')$, having no more 
than $M$ states each, with first--stage delay not
exceeding $d$, second--stage delay not exceeding $d'$, and which satisfy the
rate constraints:
\begin{eqnarray}
\sum_{i=1}^n L(u_i)&\le&nR\nonumber\\
\sum_{i=1}^n L'(v_i)&\le&n\Delta R
\end{eqnarray}
and
\begin{eqnarray}
\Delta(x^n,\calE,\calD)&\le&D_1\nonumber\\
\Delta'(x^n,\calE',\calD')&\le&D_2
\end{eqnarray}
Let $\Delta_{M,d,d'}(x^n,R,\Delta R)$ denote 
the set of distortion pairs $(D_1,D_2)$ that are
$(M,d,d')$--achievable for $x^n$. 

While the definition of $\Delta_{M,d,d'}(x^n,R,\Delta R)$ is operational,
we next define an informational achievable 
region $\Delta_{X^\ell|Y^\ell,Z^\ell}(R,\Delta R)$
(with $X^\ell$ being distributed according to 
the empirical distirbution of $\ell$--vectors)
as follows: $(D_1,D_2)\in \Delta_{X^\ell|Y^\ell,Z^\ell}(R,\Delta R)$ iff 
there exist random variables $U,V$ that satisfy the Markov relation
$(U,V)\to X^\ell\to (Y^\ell,Z^\ell)$ and functions $h$ and $h'$
such that
\begin{eqnarray}
E\rho(X^\ell,h(Y^\ell,U))&\le&\ell D_1\nonumber\\
E\rho'(X^\ell,h'(Z^\ell,U,V))&\le&\ell D_2\nonumber\\
H(U)&\le&\ell R\nonumber\\
H(V|U)&\le&\ell \Delta R
\end{eqnarray}
In the theorem below, which is an extension of Theorem \ref{thm1},
$\Delta_{X^\ell|Y^\ell,Z^\ell}(R,\Delta R)-(\delta,\delta')$ means the set
$\{(D_1-\delta,D_2-\delta'):~(D_1,D_2)\in 
\Delta_{X^\ell|Y^\ell,Z^\ell}(R,\Delta R)\}$.

\begin{theorem}:
For every positive integers $n$ and $\ell$ such that $\ell$ divides $n$:
\begin{equation}
\Delta_{M,d,d'}(x^n,R,\Delta R)\subseteq
\Delta_{X^\ell|Y^\ell,Z^\ell}\left(R+\frac{2\log M}{\ell},\Delta R+
\frac{\log M}{\ell}\right)
-\frac{1}{\ell}(\rho_{\max}d,\rho_{\max}'d').
\end{equation}
where $\rho_{\max}'\dfn\max_{x,\tx}\rho'(x,\tx)$. 
\end{theorem}

{\it Proof.}
The first stage is as before.
As for the second stage, similarly to (\ref{17}), (\ref{18}) and (\ref{20}), we have
\begin{eqnarray}
\ell \Delta R &\ge&  H(V^\ell|U^\ell,T')\nonumber\\
&\ge&  H(V^\ell,T'|U^\ell,S,S',T')\nonumber\\
&\ge& H(V^\ell,T,T'|U^\ell,S,S')-\log M
\end{eqnarray}
Also, $(U^\ell,V^\ell)\to (X^\ell,S,S',T,T')\to (Y^\ell,Z^\ell)$ is a
Markov chain. Again, since 
$(S,S',T,T')\to X^\ell\to (Y^\ell,Z^\ell)$ is also
a Markov chain, then so is 
$(U^\ell,S,S',V^\ell,T,T')\to X^\ell\to Y^\ell\to Z^\ell$.
The reconstructed output $\tX^{\ell-d'}$ 
is a function of $(U^\ell,V^\ell,T',Z^\ell)$
which is a special case of a function of 
$(U^\ell,S,S',V^\ell,T,T',Z^\ell)$ and the
distortion at the second stage is then lower 
bounded by $\frac{1}{\ell}E\rho'(X^\ell,\tX^\ell)-
\rho_{\max}'d'/\ell$ as before.
Hence, defining $U=(U^\ell,S,S')$ and $V=(V^\ell,T,T')$, 
we have found RV's $(U,V)$
and functions $h$ and $h'$ that satisfy the conditions for 
$(D_1-\rho_{\max}d/\ell,D_2-\rho_{\max}'d'/\ell)$ being in 
$\Delta_{X^\ell|Y^\ell,Z^\ell}(R+2\log M/\ell,\Delta R+
2\log M/\ell)$ provided that 
$(D_1,D_2)\in\Delta_{M,d,d'}(x^n,R,\Delta R)$.
$\Box$

The achievability is conceptually simple. Again, the first stage is as before. 
The second stage is a conditional version of 
the first where both encoder and decoder
have access to the already decoded $U$.

Finally, a few comments are in order, in addition to the comments made
at the Discussion of Section 3:
\begin{itemize}
\item [1.]
While in \cite{SM04} there is no apparent way to generalize the results to
more than two stages (unless the SI's are identical), 
here the extension is straightforward.
\item [2.]
Unlike in \cite{SM04}, here there is no need for the Markov structure $X\to Z\to Y$.
\item [3.]
The alphabet sizes required for $U$ 
and $V$ are $|\calU|\le \alpha^\ell+3$ and
$|\calV|\le \alpha^\ell\cdot |\calU|+1$.
\end{itemize}

\section*{Acknowledgement}
Neri Merhav would like to thank Tsachy Weissman for useful discussions
at the early stages of this work.

\end{document}